\documentclass[apj,iop]{emulateapj}
\usepackage{lineno}

\usepackage{natbib}
\usepackage{longtable}
\usepackage{graphicx}
\usepackage[caption=false]{subfig}
\usepackage{float}
\usepackage{epstopdf}
\usepackage{amsmath}
\usepackage{appendix}
\usepackage{hyperref}
\usepackage{booktabs}
\usepackage{scrextend}
\usepackage{xcolor}
\usepackage{CJKutf8}
\usepackage{gensymb}

\newcommand{\msun}{$M_\odot$}

\shorttitle{The AGN Fraction in Dwarf Galaxies from eROSITA}
\shortauthors{Latimer et al.}

\begin{document}

\title{The AGN Fraction in Dwarf Galaxies from eROSITA: \\ First Results and Future Prospects}

\author{Colin J. Latimer}
\affil{eXtreme Gravity Institute, Department of Physics, Montana State University, Bozeman, MT 59717, USA}
\email{colin.latimer@montana.edu}

\author{Amy E. Reines}
\affil{eXtreme Gravity Institute, Department of Physics, Montana State University, Bozeman, MT 59717, USA}

\author{Akos Bogdan}
\affil{Harvard/Smithsonian Center for Astrophysics, 60 Garden Street, Cambridge, MA 02138, USA}

\author{Ralph Kraft}
\affil{Harvard/Smithsonian Center for Astrophysics, 60 Garden Street, Cambridge, MA 02138, USA}

\begin{abstract}

Determining the fraction of nearby dwarf galaxies hosting massive black holes (BHs) can inform our understanding of the origin of ``seed" black holes at high redshift.
Here we search for signatures of accreting massive BHs in a sample of dwarf galaxies ($M_\star \le 3 \times 10^9~M_\odot$, $z \leq 0.15$) selected from the NASA-Sloan Atlas (NSA) using X-ray observations from the eROSITA Final Equatorial Depth Survey (eFEDS). On average, our search is sensitive to active galactic nuclei (AGNs) in dwarf galaxies that are accreting at $\gtrsim 1\%$ of their Eddington luminosity. Of the ${\sim}28,000$ X-ray sources in eFEDS and the $495$ dwarf galaxies in the NSA within the eFEDS footprint, we find six galaxies hosting possible active massive BHs. If the X-ray sources are indeed associated with the dwarf galaxies, the X-ray emission is above that expected from star formation, with X-ray source luminosities of $L_{0.5-8~\textrm{keV}} \sim 10^{39\textrm{-}40}$ erg s$^{-1}$. {Additionally, after accounting for chance alignments of background AGNs with dwarf galaxies, we estimate there are between 0-9 real associations between dwarf galaxies and X-ray sources in the eFEDS field at the 95\% confidence level.
From this we find an upper limit on the eFEDS-detected dwarf galaxy AGN fraction of $\le 1.8$\%}, which is
broadly consistent with similar studies at other wavelengths. We extrapolate these findings from the eFEDS sky coverage to the planned eROSITA All-Sky Survey and estimate that upon completion, the all-sky survey could yield {as many as ${\sim}1350$} AGN candidates in dwarf galaxies at low redshift.

\end{abstract}

\section{Introduction}\label{sec:intro}

Massive black holes (BHs) with $M_{\rm BH} \sim 10^{6-9} M_\odot$ reside at the center of nearly all massive galaxies. Some of these BHs are accreting at sufficient rates such that they are detectable as active galactic nuclei (AGNs). The prevalence of massive BHs and the AGN fraction in dwarf galaxies ($M_\star \lesssim 10^{9.5}$~\msun; $M_{\rm BH} \lesssim 10^5 M_\odot$) is much more uncertain, yet extremely important for our understanding of the origin of BH seeds \citep{greene20}, the role of BH feedback at low masses \citep{silk17}, and the overall demographics of massive BHs since dwarfs are the most numerous types of galaxies in the Universe.

There are various ways to search for active massive BHs in dwarf galaxies, including optical emission line diagnostics \citep[e.g.,][]{reines13,molina21}, optical variability \citep[e.g.,][]{baldassare20}, mid-IR colors \citep[e.g.,][]{hainline16,lupi20,latimer21} and radio observations \citep{mezcua19,reines20}. 
%
While all techniques suffer from various selection effects, X-ray observations are particularly well-suited to detecting BHs that are accreting at low Eddington ratios \citep{gallo08,hickox09} and/or residing in host galaxies with ongoing star formation \citep{reines16,kimbro21}.


Previous X-ray studies have been used to explore the BH occupation fraction in low mass early-type spheroidal galaxies and late-type spirals, as well as to search for AGNs in the general population of dwarf galaxies.
For example, \cite{miller15} used {\it Chandra} X-ray data from the AMUSE\footnote{AGN Multiwavelength Survey of Early-Type Galaxies}-Virgo \citep{gallo08,gallo10} and AMUSE-Field \citep{miller12} surveys to constrain the BH occupation fraction to be $>20$\% for early-type galaxies with $M_\star \sim 10^9 - 10^{10}$~\msun. A similar occupation fraction is found for late-type spirals by \citet{she17} using data in the {\it Chandra} archive. Candidate AGNs in dwarf galaxies with $M_\star \lesssim 10^{9.5}$~\msun\ have been found using the Chandra Deep Field South Survey \citep{schramm13}, archival data in the Chandra Source Catalog \citep{lemons15}, the Chandra COSMOS-Legacy Survey \citep{mezcua18} and the 3XMM catalog \citep{birchall20}.

 While much progress has been made in recent years, previous X-ray studies have been affected by incompleteness effects, small sample sizes, and are generally expensive with pointing telescopes since most of the dwarf galaxies will not have an AGN. Large-area X-ray surveys can overcome these issues and are one of the best methods to identify AGNs. In this Letter, we explore the AGN fraction in dwarf galaxies using unbiased X-ray observations from the eROSITA Final Equatorial Depth Survey \citep[eFEDS;][]{brunner21}. This work provides a sneak peek into the very near future when the eROSITA All-Sky Survey (eRASS) will be available, which could revolutionize our understanding of AGNs in dwarf galaxies.

\section{Data} \label{sec:data}

\subsection{eROSITA/eFEDS} \label{sec:xraydata}

eROSITA is the primary instrument on the Spectrum-Roentgen-Gamma (SRG) mission. It was launched into space on July 13, 2019 and five months later started a survey of the entire sky, which is predicted to finish in 2023. eROSITA is a wide-field X-ray telescope operating in the 0.2-8 keV energy range, with an average on-axis resolution of ${\sim}16\arcsec$ at 1.5 keV. It was designed to be ${\sim}25$ times more sensitive than the ROSAT all-sky survey in the 0.2-2.3 keV band. The nominal positional accuracy of eROSITA is 3\arcsec. For more details on eROSITA, see \citet{predehl21}. 

The eFEDS catalog is based off of observations taken by eROSITA during its performance verification phase ahead of the all-sky survey. eFEDS has a sky footprint of ${\sim}140$ deg$^2$ (1/300th of the sky), which is broken up into four ${\sim}35$ deg$^2$ rectangular, semi-adjacent sub-fields of $4.2\degree \times 7.0\degree$ \citep[for a visualization, see Figure 1 in][]{brunner21}. This observational strategy was designed to provide uniform exposure over this field and to be ${\sim}50$\% deeper than is expected for the future eROSITA all-sky survey. The eFEDS catalog contains $27,910$ X-ray sources detected in the 0.2-2.3 keV band, with a nominal (point source) flux limit of ${\sim}7 \times 10^{-15}$ erg cm$^{-2}$ s$^{-1}$ in the $0.5-2$~keV band $-$ at this level, the fraction of spurious contaminating sources is expected to be 6.4\%. The completeness is given as 96\% at a $0.5-2$~keV band flux limit of $10^{-14}$ erg cm$^{-2}$ s$^{-1}$, and down to 65\% at a flux limit of  $4 \times 10^{-15}$ erg cm$^{-2}$ s$^{-1}$.

The eFEDS catalog has a source density larger than the ROSAT All-Sky survey \citep{boller16} by a factor of 70, and has a quite uniform source density across its coverage. Astrometric corrections were found by matching eFEDS sources to the catalog of Gaia-unWISE AGN candidates from \cite{shu19}, as described in \cite{brunner21}. Additionally, \cite{brunner21} compare the eFEDS source fluxes to those of sources in the same footprint appearing in the XMM-ATLAS survey \citep{ranalli15}, and find no evidence for strong systematic offsets.

\subsection{Sample of Dwarf Galaxies} \label{sec:nsadata}

We draw our parent sample of dwarf galaxies from the NSA (v1\_0\_1), a catalog of local ($z \leq 0.15$) galaxies and associated parameters derived via images from the Sloan Digital Sky Survey (SDSS) and the \textit{Galaxy Evolution Explorer (GALEX)}. Galaxy stellar masses are calculated using \texttt{kcorrect} v4\_2 \citep{blanton07}, and are given in units of \msun $h^{-2}$. We adopt $h=0.73$, and use the catalog values derived from elliptical Petrosian photometry. We filter the NSA for dwarf galaxies by excluding any galaxies with masses above $3 \times 10^{9}$ \msun. This results in a parent sample of 63,582 dwarf galaxies.

While we use this entire parent sample for matching dwarf galaxies to eFEDS sources (Section \ref{sec:galmatch}), we note that the eFEDS sky coverage is significantly smaller than that of the NSA; 140 deg$^{2}$ \citep{brunner21} vs. 14,555 square degrees\footnote{\url{https://www.sdss.org/dr13/scope/}}. To estimate how many dwarf galaxies (in the NSA catalog) reside within the eFEDS footprint, we first approximate the eFEDs sky coverage by using a bounding box, determined by the maximum and minimum X-ray source R.A. and Declination values as reported in the eFEDS catalog. We then search the NSA for dwarf galaxies lying within this area, resulting in 495 dwarf galaxies (an upper limit, given our use of a bounding box).

We also note that, while the NSA sky footprint overlays nearly the entire eFEDS coverage, there are a few small parts of the southern-most regions of eFEDS that may lie outside the NSA coverage \citep[see Figure 1 in][]{brunner21}; however, as the overwhelming majority of eFEDS is covered, we expect this effect to be negligible.

\section{Analysis and Results} \label{sec:analysis}

\subsection{Dwarf Galaxies with X-ray Detections} \label{sec:galmatch}

\begin{figure*}
\captionsetup[subfigure]{labelformat=empty}
{\includegraphics[width=0.33\textwidth]{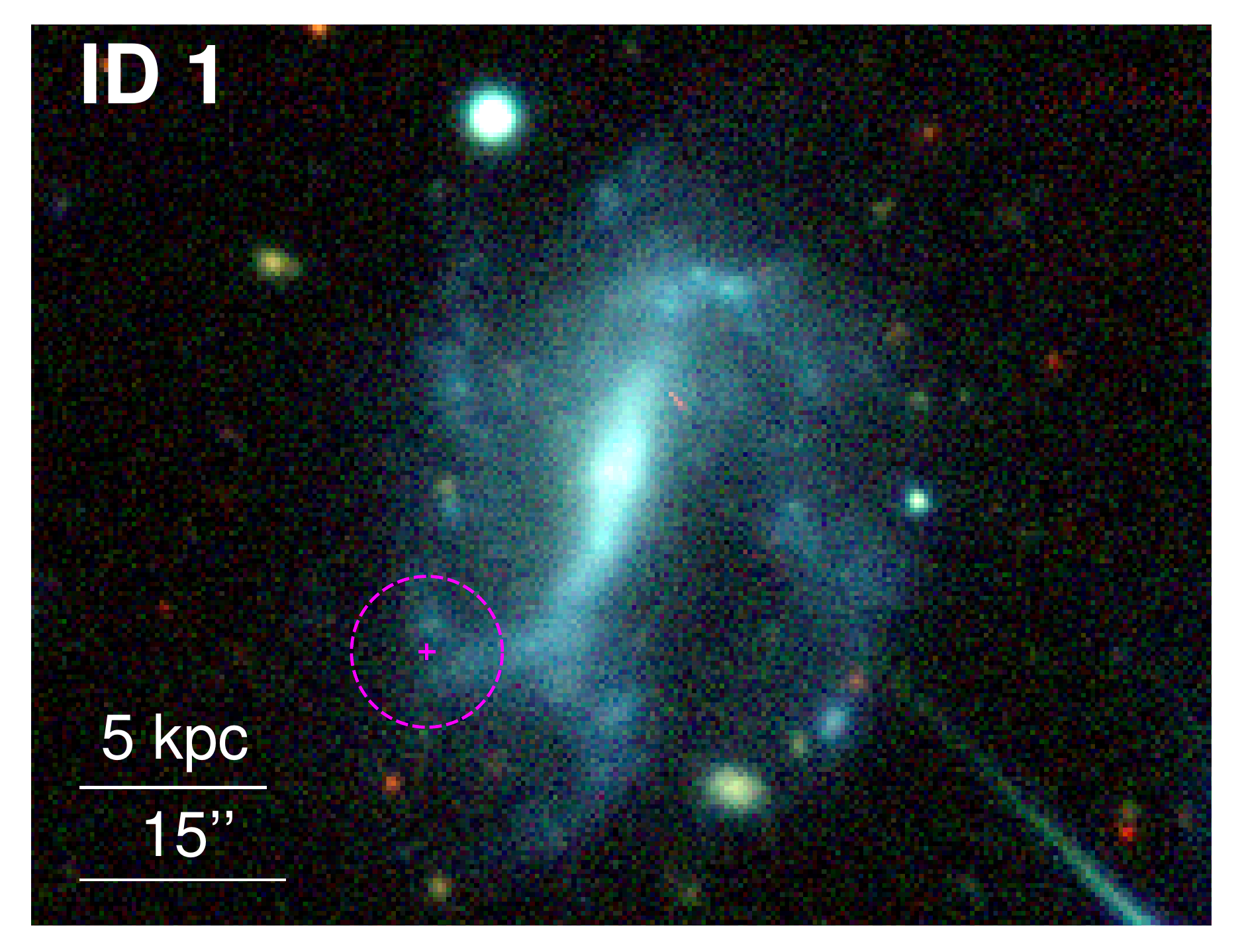}}\hfill
{\includegraphics[width=0.33\textwidth]{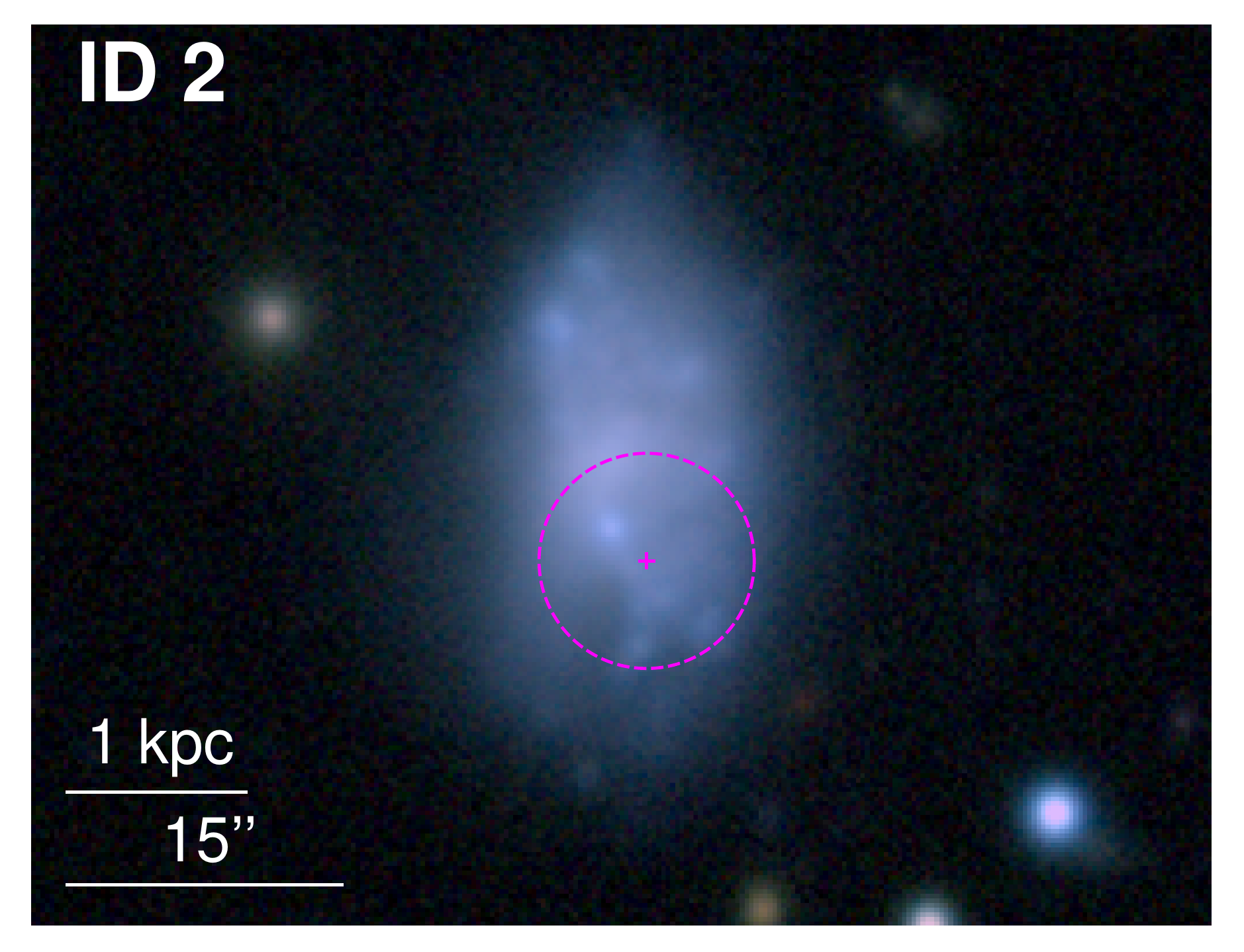}}\hfill
{\includegraphics[width=0.33\textwidth]{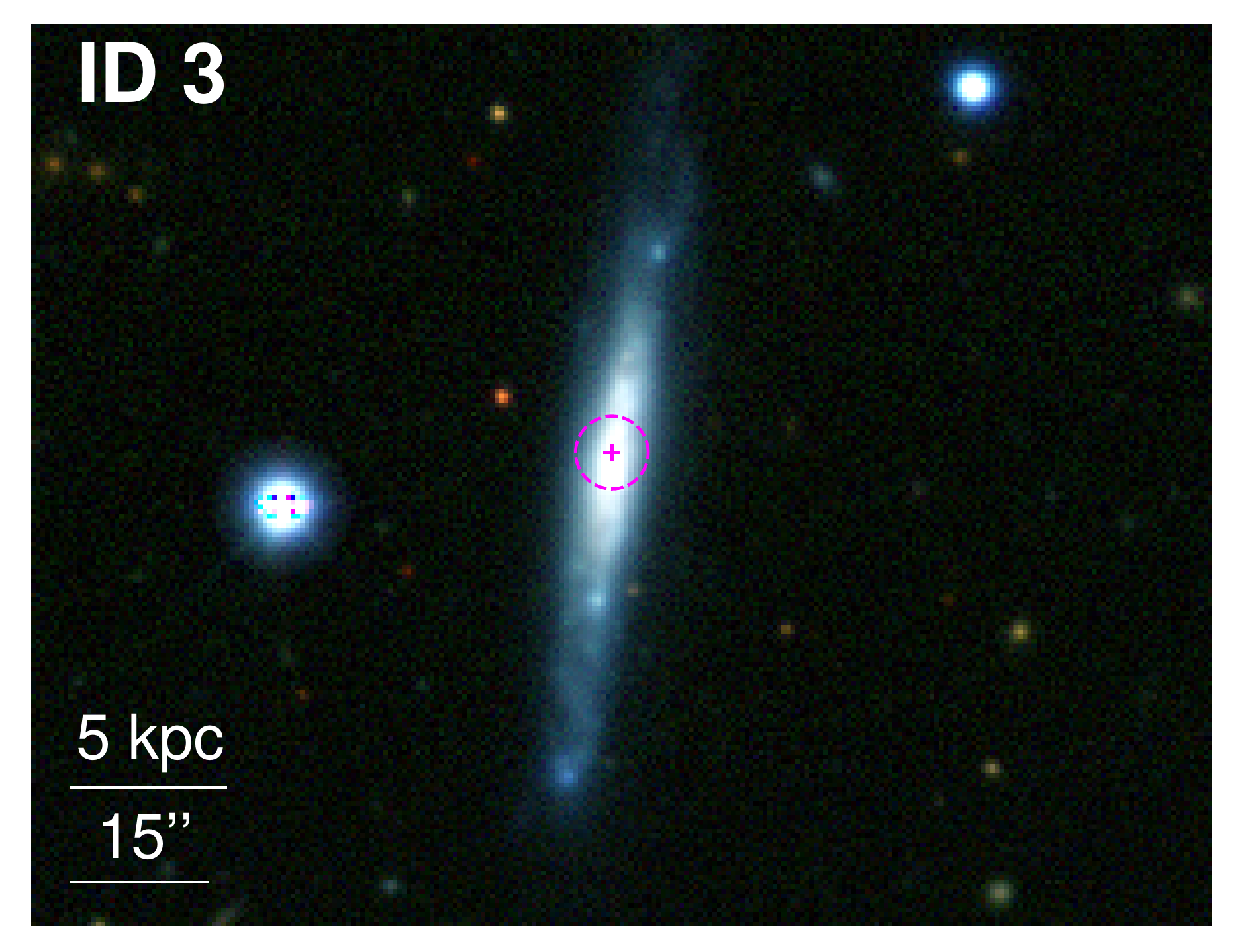}}\hfill

{\includegraphics[width=0.33\textwidth]{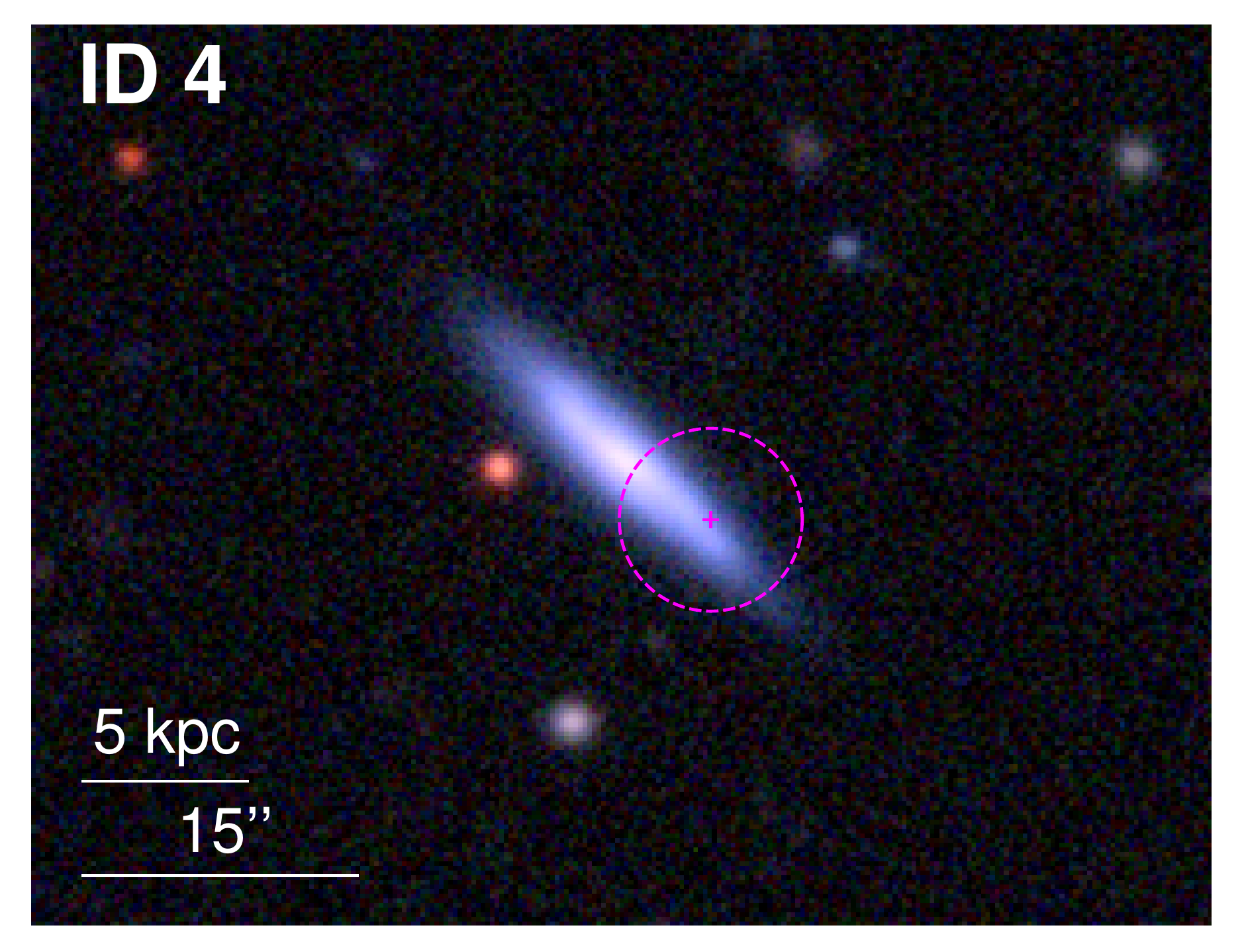}}\hfill
{\includegraphics[width=0.33\textwidth]{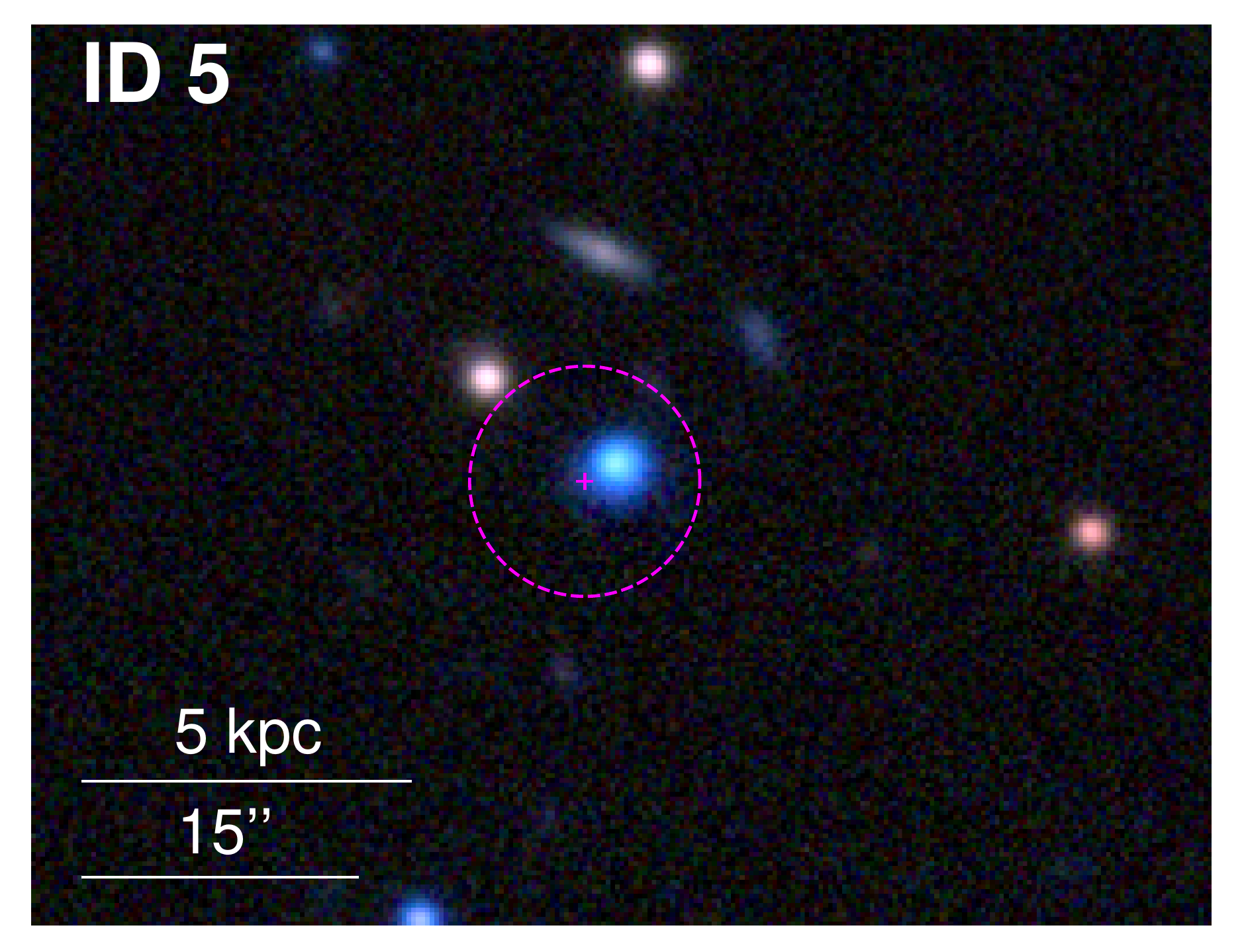}}\hfill
{\includegraphics[width=0.33\textwidth]{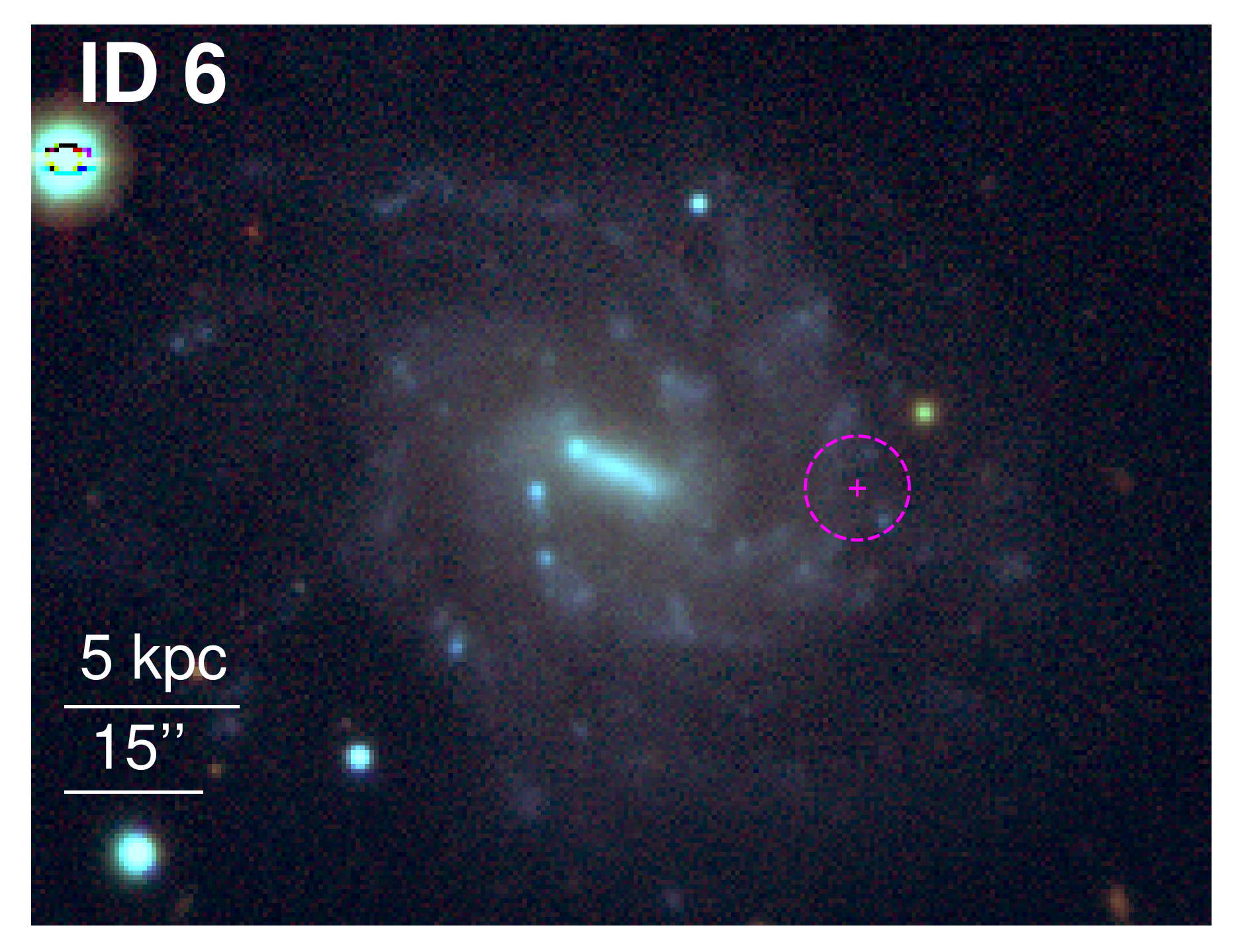}}\hfill

\caption{Three-color images of our galaxies retrieved from the Dark Energy Camera Legacy Survey \citep[DECaLS;][]{dey19}, with red, green, and blue showing the $z$, $r$, and $g$ bands, respectively. We overlay the positions and the corrected combined positional uncertainty (via the \texttt{RADEC\_ERR\_CORR} parameter from the eFEDS main catalog) in magenta.
} 
\label{fig:rgb}
\end{figure*}

To find dwarf galaxies with coincident X-ray detections, we first cross-match X-ray sources from the eFEDS main catalog\footnote{\url{https://erosita.mpe.mpg.de/edr/eROSITAObservations/Catalogues/}} with the NSA sample of 495 dwarf galaxies that lie within the eFEDS footprint. We include matches 
in which the separation between the X-ray source and the galaxy optical center is less than {two} Petrosian 50\% light radii ({2}$r_{50}$). This results in {six} galaxies matching to {six} X-ray sources {(one per galaxy; see Figure \ref{fig:rgb})}. {We 
check to ensure that our matched galaxies are indeed dwarfs by determining if they have reliable redshifts and ensuring that the absolute magnitudes are not anomalously bright. All six galaxies pass these tests.}




Our final sample consists of six dwarf galaxies, each with a single associated eFEDS X-ray point source. {We check these X-ray sources against the ROSAT all-sky survey source catalog \citep{boller16}, the XMM-ATLAS catalog \citep{ranalli15}, and the Chandra Source Catalog 2.0\footnote{\url{https://cxc.cfa.harvard.edu/csc/}}, finding no matches.} The galaxy properties of our sample are summarized in Table \ref{tab:sample}, and three-color optical images with X-ray source positions are shown in Figure \ref{fig:rgb}. Only one galaxy has an X-ray source clearly associated with the nucleus (ID 3). The remaining five X-ray sources have centroids that are offset from the optical centers of the galaxies by {1\farcs8}-25\farcs8, although three of these {(IDs 2,4,5)} are marginally consistent with the optical centers within the X-ray positional uncertainties (see Figure \ref{fig:rgb}). While off-nuclear AGNs are possible \citep{bellovary19,reines20}, we also consider chance alignments between the dwarf galaxies and background X-ray sources.


\begin{deluxetable*}{cccccccccc}
\tabletypesize{\footnotesize}
\tablecaption{Sample}
\tablewidth{0pt}
\tablehead{
\colhead{ID} & \colhead{NSAID} & \colhead{R.A.} & \colhead{Decl.} & \colhead{$N_{\rm H}$} & \colhead {z} & \colhead{$r_{50}$} & \colhead{Distance} & \colhead{log $M_\star$/\msun} & \colhead{SFR}  \\
\colhead{ } & \colhead{ }   & \colhead{(deg)} & \colhead{(deg)}  & \colhead{($10^{20}$ cm$^{-2}$)} & \colhead{ } & \colhead{(kpc)}& \colhead{(Mpc)} & \colhead{ } & \colhead{(\msun~yr$^{-1}$)}  \\
\colhead{(1)} & \colhead{(2)} & \colhead{(3)} & \colhead{(4)} & \colhead{(5)} & \colhead{(6)}  & \colhead{(7)} & \colhead{(8)} & \colhead{(9)} & \colhead{(10)} }
\startdata
1 & 82162  & 140.942967 & 2.753302    & 3.58  & 0.0177 & 8.24 & 73  & 9.08 & 0.16  \\
2 & 623354 & 145.180051 & 3.958864    & 3.64  & 0.0051 & 0.91 & 21  & 8.40 & 0.04  \\
3 & 648474 & 141.572049 & 3.134800    & 3.82  & 0.0149 & 3.24 & 61  & 9.44 & 0.27  \\
4 & 56620  & 134.438656 & $-0.356626$ & 3.28  & 0.0277 & 2.69 & 114 & 9.09 & 0.32  \\
5 & 623313 & 143.913363 & $-1.261538$ & 3.38  & 0.0141 & 0.30 & 58  & 6.87 & 0.07  \\
6 & 647996 & 133.967782 & 2.524284    & 3.89  & 0.0132 & 6.35 & 54  & 9.43 & 0.34 
\enddata
\tablecomments{Column 1: Identification number used in this paper.
Column 2: NSAIDs from  v1\_0\_1 of the NSA.
Column 3: right ascension of the galaxy.
Column 4: declination of the galaxy.
Column 5: galactic neutral hydrogen column density \citep{dickey90}\footnote{Retrieved via \url{https://cxc.harvard.edu/toolkit/colden.jsp}}.
Column 6: redshift, specifically the \texttt{zdist} parameter from the NSA.
Column 7: Petrosion 50\% light radius.
Column 8: galaxy distance.
Column 9: log galaxy stellar mass.
Column 10: estimated SFRs from {\it GALEX} and {\it WISE} data (see Section \ref{sec:xraydets}).
The values given in columns 6-9 are from the NSA and we assume $h=0.73$.}
\label{tab:sample}
\end{deluxetable*}

\subsection{Chance Alignments with Background Sources} \label{sec:valsrc}

\begin{figure}[h!]
\centering
\includegraphics[width=0.48\textwidth]{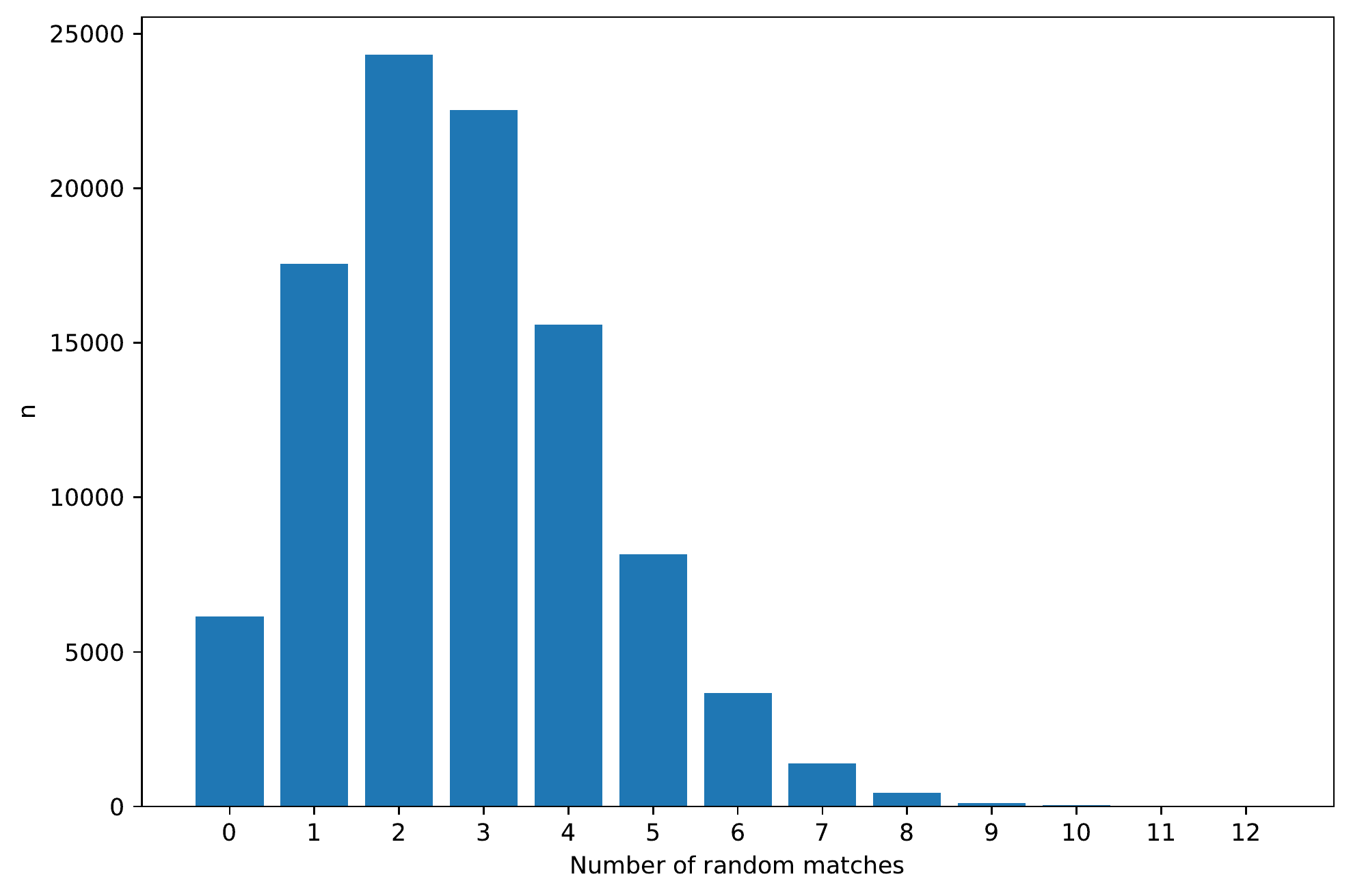}
\caption{The results of matching $10^5$ randomly generated coordinate sets (in the eFEDS field) to the eFEDS X-ray sources.
}
\label{fig:bgsrcs}
\end{figure}

Due to the large number of eFEDS sources (${\sim}28,000$), some of the galaxies in our sample may have matched to X-ray sources purely due to spatial coincidence. Here we estimate how many matches we would expect to find due to this phenomenon.
Following the same general method as in \cite{kovacs20}, we start with our estimation of how many (NSA-detected) dwarf galaxies lie in the eFEDS footprint: 495 galaxies (see Section \ref{sec:nsadata}).
Next, we perform Monte Carlo simulations to estimate how many of these 495 dwarf galaxies may have coincidentally matched with one or more eFEDS sources. We randomly generate 495 coordinate pairs within the bounding box of the eFEDS footprint and match these random coordinates to the eFEDS X-ray sources. We use the same matching criteria as we used for our sample selection in Section \ref{sec:galmatch} (i.e., the separation between the eFEDS and `galaxy' sources be less than {$2r_{50}$}). As our randomly generated coordinates do not come with attached half-light radii, we instead take the $r_{50}$ values from the actual dwarf galaxies (via the NSA) and randomly assign one to each generated coordinate. Note that of the 495 dwarf galaxies, five had bad $r_{50}$ data; for these, we substituted the median $r_{50}$ value of ${\sim}3\farcs7$ (we also obtained near-identical results when instead using the mean $r_{50}$ value of ${\sim}4\farcs6$, or simply excluding those five galaxies/coordinates altogether). We record the number of coordinate pairs that match to at least one eFEDS X-ray source, and then repeat this process $10^5$ times. 
The results are shown in the histogram in Figure \ref{fig:bgsrcs}. 
The distribution of random matches in our Monte Carlo simulations has a mean and median of {${\sim}3$}. 

{We estimate the 95\% confidence interval for the number of dwarf galaxies with X-ray sources in eFEDS following the \cite{kraft91} Bayesian formalism for Poisson-distributed data with low counts in the presence of a background. We have six `counts' (the six detected matches in our sample) with a background of three counts (from chance alignment), which results in lower and upper limits of $(0,8.90)$. From this we conclude that there are between 0-9 actual X-ray matches to dwarf galaxies in the eFEDS field. Note that these matches represent dwarf galaxies with an associated X-ray source, but they are not necessarily AGNs (see Section \ref{sec:xraydets} below).} 

\subsection{Origin of the X-ray Emission in Dwarf Galaxies} \label{sec:xraydets}

Here we discuss the origin of the X-ray emission under the assumption that the X-ray sources are indeed associated with the matched dwarf galaxies (although see Section \ref{sec:valsrc} above).
To get X-ray source properties, we re-analyze the eFEDS images\footnote{retrieved from \url{https://erosita.mpe.mpg.de/edr/eROSITAObservations/}}. We filter the images to the $0.5-2$~keV energy band to cut down on noise for our measurements. To extract counts, we use circular source regions with radii of 20\arcsec and accompanying background annuli regions with inner and outer radii of 30\arcsec and 80\arcsec, respectively (except for ID 1 $-$ it had other sources nearby, so we used a background region with 30\arcsec and 55\arcsec inner and outer radii). We convert our measured net counts to $0.5-2$~keV fluxes using the $0.5-2$~keV band Energy Conversion Factor (ECF) from \cite{brunner21} of $1.185 \times 10^{12}$ cm$^2$ erg$^{-1}$, which was found assuming an absorbed power-law with a slope of 2.0 and a Galactic absorption of $3 \times 10^{20}$ cm$^{-2}$. 

To get flux values in the $0.5-8$~keV band, we apply a correction factor of ${\sim}0.32$ dex (a factor of ${\sim}2.07$), calculated via the \textit{Chandra}-based Portable, Interactive Multi-Mission Simulator (PIMMS)\footnote{\url{https://cxc.harvard.edu/toolkit/pimms.jsp}}, assuming a photon index of $\Gamma=2$ and an absorption of $N_{\rm H}= 3 \times 10^{20}$ cm$^{-2}$.
Note also that our sources have differing spectra (some harder than the others) so while we are consistent in our methodology, the accuracy of this one-size-fits-all extrapolation approach may vary from source to source.
Our resulting $0.5-8$~keV log luminosities range between log $L_{0.5-8~\textrm{keV}} \sim 39.1\textrm{-}40.0$ erg s$^{-1}$ (Table \ref{tab:xray}). These values are similar to those of other AGNs in dwarf galaxies \citep[e.g.,][]{reines14,baldassare17, mezcua18}.

\begin{deluxetable*}{cccccc}
\tabletypesize{\footnotesize}
\tablecaption{X-ray sources}
\tablewidth{0pt}
\tablehead{
\colhead{ID} & \colhead{R.A.} & \colhead{Decl.} & \colhead{Sep.} & \colhead{$F_{0.5-8~\textrm{keV}}$}  & \colhead{log $L_{0.5-8~\textrm{keV}}$} \\
\colhead{ } & \colhead{(deg)} & \colhead{(deg)} & \colhead{(kpc)} & \colhead{($10^{-15}$ erg s$^{-1}$ cm$^{-2}$)} & \colhead{(erg s$^{-1}$)} \\
\colhead{(1)} & \colhead{(2)} & \colhead{(3)} & \colhead{(4)} & \colhead{(5)} & \colhead{(6)}  }
\startdata
1  & 140.9471979 & 2.7490836    & 7.59 & 3.86     & 39.4        \\
2  & 145.1794807 & 3.9583302    & 0.29 & 26.21    & 39.1        \\
3  & 141.5719005 & 3.1352167    & 0.47 & 12.74    & 39.8        \\
4  & 134.4373197 & $-0.3575533$ & 3.23 & 6.97     & 40.0        \\
5  & 143.9138111 & $-1.2617757$ & 0.51 & 21.34    & 39.9        \\
6  & 133.9606284 & 2.5236514    & 6.80 & 12.19    & 39.6
\enddata
\tablecomments{Column 1: galaxy ID associated with the X-ray source.
Column 2: right ascension of the X-ray source.
Column 3: declination of the X-ray source.
Column 4: separation between galaxy center and X-ray source position.
Column 5: Flux in $0.5-8$~keV band, in erg s$^{-1}$ cm$^{-2}$.
Column 6: Log luminosity in $0.5-8$~keV band, in erg s$^{-1}$.
}
\label{tab:xray}
\end{deluxetable*}



We next compare our observed X-ray source luminosities to the expected galaxy-wide contribution from 
X-ray binaries (XRBs). While enhanced X-ray emission could indicate the presence of an AGN, it is not a required condition for one. 
The expected X-ray luminosity from low-mass XRBs scales with stellar mass \citep{gilfanov04}, and with star-formation rate (SFR) for high-mass XRBs \citep{grimm03,mineo12}. We use the relation of \cite{lehmer10}, which accounts for both: $L_{2-10~\textrm{keV}}^{\rm XRB}({\rm erg~s}^{-1}) = \alpha M_{*} + \beta \textrm{SFR}$ with $\alpha = (9.05 \pm 0.37) \times 10^{28}$ erg s$^{-1}$ \msun$^{-1}$ and $\beta = (1.62 \pm 0.22) \times 10^{39}$ erg s$^{-1}$ (\msun~yr$^{-1}$)$^{-1}$. The relation has a scatter of 0.34 dex.
The \cite{lehmer10} relation uses $2-10$~keV X-ray luminosities; we convert our $0.5-8$~keV luminosities to this band using a correction factor of $-0.22$ dex (a factor of ${\sim}1.7$), found using PIMMS in the same manner as before.

We estimate SFRs of the galaxies using mid-infrared (IR; 25 $\mu$m) and far-UV (FUV; 1528 \AA) luminosities via the relations 
\begin{equation} \label{eq:sfrs}
\begin{gathered}
 \textrm{log SFR}(\textrm{\msun~yr}^{-1}) = \textrm{log }L\textrm{(FUV)}_{\textrm{corr}} - 43.35 \\
 L \textrm{(FUV)}_{\textrm{corr}} = L\textrm{(FUV)}_{\textrm{obs}} + 3.89 L\textrm{(25 $\mu$m)}
\end{gathered}
\end{equation}
\citep{kennicutt12,hao11}. The uncertainty in the resulting SFRs is ${\sim}0.13$ dex.
We retrieve FUV magnitudes from \textit{GALEX} via the NSA. We use 22 $\mu$m magnitudes from the \textit{Wide-field Infrared Survey Explorer (WISE)} in lieu of 25 $\mu$m luminosities from the \textit{Infrared Astronomical Satellite (IRAS)} since none of our galaxies were detected by \textit{IRAS} and all six of our galaxies had \textit{WISE} detections.  Moreover, the ratio between 25 $\mu$m and 22 $\mu$m flux densities is expected to be of order one \citep{jarrett13}, making 22 $\mu$m observations a reasonable proxy. The resulting SFRs span a range of 0.04-0.34 \msun~yr$^{-1}$, with a median of 0.21 \msun~yr$^{-1}$ (Table \ref{tab:sample}).

\begin{figure}[h!]
    \centering
    \includegraphics[width=0.5\textwidth]{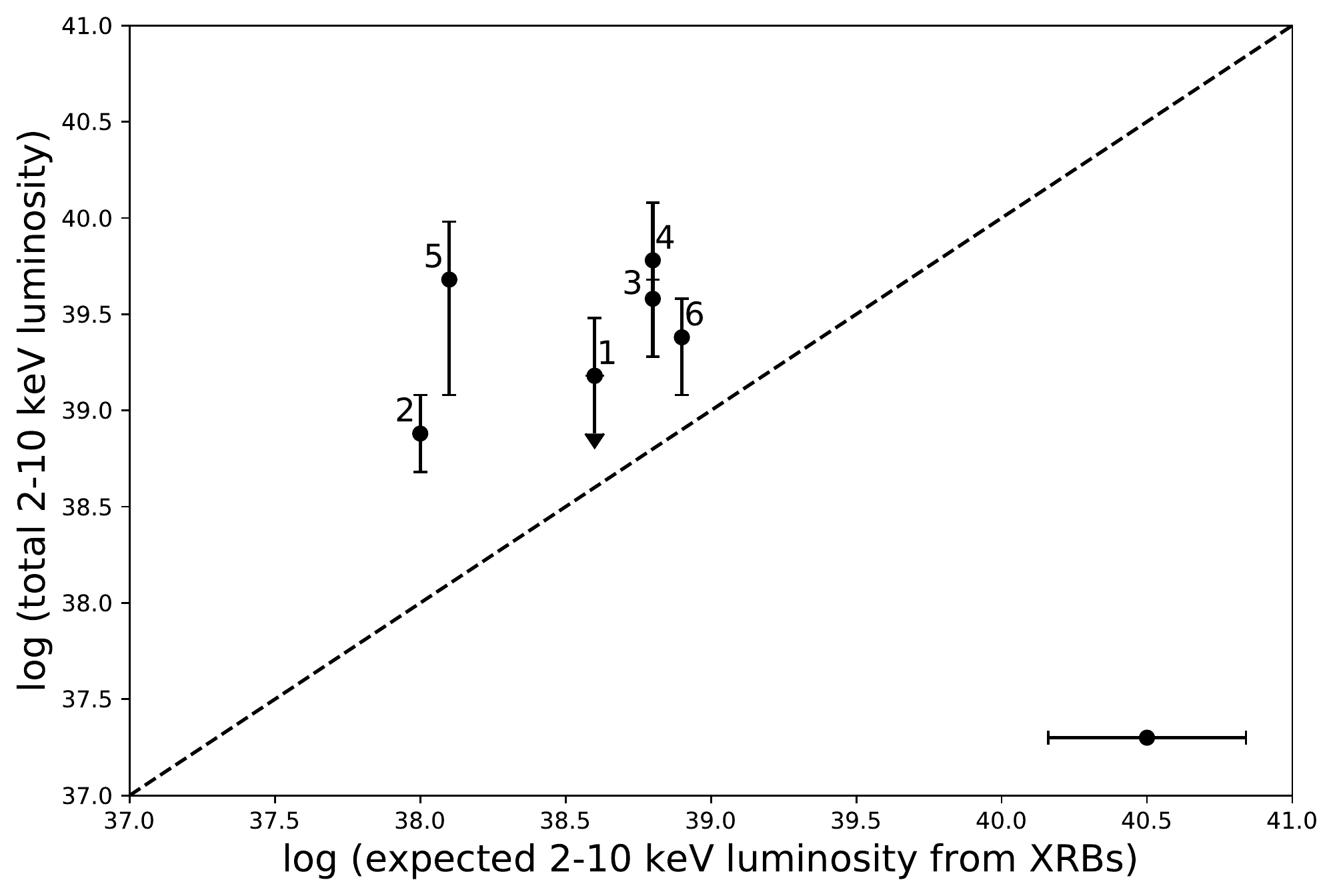}
    \caption{Total observed X-ray luminosities of the galaxies in our sample vs. what we would expect to see from XRBs via the relation of \cite{lehmer10}. The dashed line represents the one-to-one relation, and the scatter of 0.34 dex in the expected XRB luminosities is shown in black in the lower right corner. The error bars shown are calculated using the 1$\sigma$ error approximation from \cite{gehrels86}.
    Note also that the lower error on ID 1 was consistent with no detection; this is signified by the arrow for the lower error bar.}
    \label{fig:xrbplot}
\end{figure}

We plot the observed X-ray luminosities of our sources vs.\ the expected cumulative X-ray luminosity from XRBs in the host galaxies in Figure \ref{fig:xrbplot}.  All of the galaxies have moderately significant enhanced X-ray emission at the level of {0.5-1.7} dex. 
Given the relatively course angular resolution of the X-ray data (${\sim}16\arcsec$ \citealt{predehl21}), we also consider the expected galaxy-wide luminosities from diffuse X-ray emission; {we use the relation from \cite{mineo12_2}, which gives the expected diffuse X-ray emission as a function of host galaxy SFR. The resulting luminosities} (in the $0.5-2$~keV band) are consistently lower than the expected XRB luminosities (in the $2-10$~keV band) by {0.3-0.5 dex (factors of 2-3)}. 
In sum, the observed X-ray luminosities are suggestive of AGNs in the host galaxies but higher-resolution X-ray observations (e.g., with \textit{Chandra}) are needed to provide a more accurate description of the emission from these galaxies.





We also consider the possibility of ultra-luminous X-ray (ULX) sources in our sample of six dwarf galaxies. 
The observed X-ray luminosities of our detected sources are consistent with ULXs, which are defined as off-nuclear X-ray sources having $L_X > 10^{39}$ erg s$^{-1}$  \citep[see][for a review]{kaaret17}. 
{We estimate the expected number of ULXs in our sample of six galaxies, which scales with SFR and metallicity \citep{prestwich13}. We used the combined SFR of our sample (${\sim}1.2$ \msun~yr$^{-1}$). Half of our sample (IDs 2-4) had line fluxes in the NSA which we could use to estimate metallicity via the relation from \cite{pettini04}; these three fell into the intermediate or high metallicity bins used in Table 10 of \cite{prestwich13}. We make the assumption that the remaining three galaxies in our sample will fall into the same metallicity bins; this results in $<1$ expected ULX in our sample (using either high or intermediate metallicity). As a consistency check, we also use the results of \cite{swartz04} and \cite{mapelli10}. \cite{swartz04} relate the expected number of ULXs in a sample to SFR and galaxy mass independently $-$ these both result in $<1$ expected ULX in our sample as well. Using the results of \cite{mapelli10}, which are SFR-based, we find an estimated ${\sim}1.5$ ULXs in our sample of six galaxies. This suggests that ${\sim}1$ of our X-ray sources could be a ULX. If we incorporate this possibility into the confidence intervals as estimated in Section \ref{sec:valsrc}, this would translate to a background of four counts (instead of three), giving us 95\% confidence intervals of $(0,8.05)$.}

\section{Conclusions and Discussion}

\subsection{eROSITA Constraints on the AGN Fraction in Low-redshift Dwarf Galaxies}

{From matching dwarf galaxies in the NSA to X-ray sources in the recently released eFEDS catalog, and accounting for chance alignments with background AGNs, we find that there could be between 0-9 real associations between dwarf galaxies and X-ray sources. This puts an upper limit of $\leq1.8$\% on the eFEDS-selected dwarf galaxy AGN fraction.}
This AGN fraction is broadly consistent with findings from studies at optical \citep[${\sim}0.5$\%;][]{reines13} and mid-IR \citep[${\sim}0.4$\%;][]{lupi20} wavelengths. 

There are a number of caveats to consider when interpreting these results. First, we are only exploring relatively nearby ($z \leq 0.15$) and bright dwarf galaxies that are contained in the NSA. There are also limits on the luminosities of AGNs that can be detected with a flux-limited X-ray survey as discussed below. Heavily obscured AGNs could also remain undetected. 


\subsection{Detection limits}

Here we explore detection limits for finding AGNs in dwarf galaxies with eROSITA/eFEDS.
We first estimate the minimum detectable $0.5-8$~keV luminosity of an X-ray source in our parent sample of dwarf galaxies. We start with the minimum detectable flux of ${\sim}7 \times 10^{-15}$ erg cm$^{-2}$ s$^{-1}$ in the $0.5-2$~keV band \citep{brunner21}. We convert this to a $0.5-8$~keV flux using PIMMS, in the same manner as in Section \ref{sec:xraydets}. We use the median distance of the dwarf galaxies in the NSA (${\sim}138$ Mpc) 
to convert to a luminosity, resulting in an estimated minimum luminosity of $L_{0.5-8~\textrm{keV}} \sim 10^{40.5}$ erg s$^{-1}$. 
Note however that the six galaxies in our sample all have distances lower than the median distance of dwarfs in the NSA. If we instead calculate the minimum detectable luminosity using the smallest distance in our sample (${\sim}21$ Mpc; Table \ref{tab:sample}), we find $L_{0.5-8~\textrm{keV}} \sim 10^{38.9}$ erg s$^{-1}$, which is consistent with our results given in Table \ref{tab:xray}. 



The Eddington ratio of an active BH is 
given by
\begin{equation} \label{eq:fedd}
f_{\rm Edd} = (\kappa L_{\rm X})/(L_{\rm Edd})
\end{equation}
where $\kappa$ is the $2-10$~keV bolometric correction, $L_{\rm X}$ is the X-ray luminosity of the BH, and $L_{\rm Edd}$ is the Eddington luminosity of the BH, which is found via
\begin{equation} \label{eq:Ledd}
L_{\textrm{Edd}} \approx 1.26 \times 10^{38}~M_{\textrm{BH}}/\textrm{\msun} \textrm{ erg s}^{-1}.
\end{equation}
We convert the minimum detectable flux using PIMMS in the same manner as before, this time from the $0.5-2$~keV band to the $2-10$~keV band, with a resulting luminosity of $L_{\rm 2-10~keV} \sim 10^{40.2}$ erg s$^{-1}$. Assuming a typical BH mass of ${\sim}10^5$ \msun\ for a dwarf galaxy \citep{reines13} and $\kappa = 20$ \citep{vasudevan09}, the resulting Eddington ratio is ${\sim} 0.04$. If we instead use the minimum luminosity found using the smallest distance in our sample, we find $L_{\rm 2-10~keV} \sim 10^{38.7}$ erg s$^{-1}$, and an Eddington fraction of ${\sim} 0.001$.

\subsection{Future Prospects with eROSITA}

Despite the small sample presented here based on the limited eFEDS catalog, the eROSITA All-Sky Survey (eRASS) is poised to be a boon for future searches and studies of AGNs in dwarf galaxies. The eFEDs catalog covers an area on the sky of only ${\sim}140$ deg$^2$ \citep{brunner21}, roughly $1/300$th of the sky. Extrapolating the number of sources in eFEDS to the entire sky, eRASS could result in as many as ${\sim}8$ million X-ray sources. However, this number is likely to be an upper limit since eFEDS is about 50\% deeper than is expected for eRASS \citep{brunner21} and we will only have access to (the German) half of the sky. Assuming a correction factor of ${\sim}1/2$, this still leaves us with as many as ${\sim}4$ million potential new X-ray sources (over ten times larger than the Chandra Source Catalog Version 2.0). Additionally, these sources will be an unbiased sample as the observation is all-sky with no pre-selection of targets.

Matching the eFEDS catalog to dwarf galaxies in the NSA, {and accounting for chance alignments with background AGNs,} yields {0-9} AGN candidates. Extrapolating these results to the completed eRASS and accounting for the correction factor of 1/2 (see above) suggests the number of future dwarf galaxy AGN candidates at low redshift {could be as many as ${\sim}$1350}, providing an ample and exciting new sample to explore.





\acknowledgements
AER acknowledges support for this work provided by NASA through EPSCoR grant number 80NSSC20M0231. {ÁB and RK were supported by NASA contract NAS8-03060.}
This work is based on data from eROSITA, the soft X-ray instrument aboard SRG, a joint Russian-German science mission supported by the Russian Space Agency (Roskosmos), in the interests of the Russian Academy of Sciences represented by its Space Research Institute (IKI), and the Deutsches Zentrum für Luft- und Raumfahrt (DLR). The SRG spacecraft was built by Lavochkin Association (NPOL) and its subcontractors, and is operated by NPOL with support from the Max Planck Institute for Extraterrestrial Physics (MPE). The development and construction of the eROSITA X-ray instrument was led by MPE, with contributions from the Dr. Karl Remeis Observatory Bamberg \& ECAP (FAU Erlangen-Nuernberg), the University of Hamburg Observatory, the Leibniz Institute for Astrophysics Potsdam (AIP), and the Institute for Astronomy and Astrophysics of the University of Tübingen, with the support of DLR and the Max Planck Society. The Argelander Institute for Astronomy of the University of Bonn and the Ludwig Maximilians Universität Munich also participated in the science preparation for eROSITA.

\bibliography{ref}

\end{document}